# Anomalous diffusion as a stochastic component in the dynamics of complex processes

Serge F. Timashev,[1,2] Yuriy S. Polyakov,[3] Pavel I. Misurkin,[4] and Sergey G. Lakeev[2]

[1]*Institute of Laser and Information Technologies, Russian Academy of Sciences, Troitsk, Pionerskaya str. 2, Moscow Region, 142190, Russia*

[2]*Karpov Institute of Physical Chemistry, Ul. Vorontsovo pole 10, Moscow 103064, Russia*

[3]*USPolyResearch, Ashland, PA 17921, USA*

[4]*Semenov Institute of Chemical Physics, Russian Academy of Sciences, Ul. Kosygina 4, Moscow 19991, Russia*

We propose an interpolation expression using the difference moment (Kolmogorov transient structural function) of the second order as the average characteristic of displacements for identifying the anomalous diffusion in complex processes when the stochastic[1] dynamics of the system under study reaches a steady state (large time intervals). Our procedure based on this expression for identifying anomalous diffusion and calculating its parameters in complex processes is applied to the analysis of the dynamics of blinking fluorescence of quantum dots, X-ray emission from accreting objects, fluid velocity in Rayleigh-Bénard convection, and geoelectrical signal for a seismic area. For all four examples, the proposed interpolation is able to adequately describe the stochastic part of the experimental difference moment, which implies that anomalous diffusion manifests itself in these complex processes. The results of this study make it possible to broaden the range of complex natural processes in which anomalous diffusion can be identified.

---

[1] The term "stochastic" in this paper refers to random variability in the signals of complex systems characterized by nonlinear interactions, dissipation, and inertia.

## I. INTRODUCTION

Anomalous diffusion [1-12] is a "random walk" process, for which the mean square displacement of system states $V(t)$, varying during time $t$ over the whole set of possible states, ($-\infty < V < \infty$), from the mean value can be written as

$$\left\langle \left[\Delta V(t)\right]^2 \right\rangle_{pdf} = 2Dt_0 \left(t/t_0\right)^{2H_1}. \tag{1}$$

Here, $D$ is the diffusion coefficient; $t_0$ is the characteristic time; $H_1$ is the Hurst constant; the averaging, denoted by the symbol $\langle\ldots\rangle_{pdf}$, is effected by introducing the probability density function $W(V, t)$, which accounts for the probability of the system state being within the given interval of states at a specific moment $t$. It is assumed that the system was found in the vicinity of the $V = 0$ state (point) at the initial time $t = 0$. Fickian diffusion ($H_1 = 0.5$) corresponds to the random walks of system states characterized by some characteristic scale $\delta V$ of the values of elementary jumps, which are associated with the system transfers between adjacent states, and by the characteristic residence time $\delta \tau$ for every state. However, if these random walks of states stochastically alternate with the jumps having anomalous values higher than $\delta V$ at the same characteristic residence times $\delta \tau$ for the given state, the so-called superdiffusion (Lévy diffusion or Lévy flights), for which $H_1 > 0.5$, can occur. If the random walks stochastically alternate with the jumps having anomalously long times of residence in some states ("stability islands" [4]), which are much larger than $\delta \tau$, for the same values of jumps $\delta V$, the so-called subdiffusion, for which $H_1 < 0.5$, can occur.

Anomalous (from Fick's viewpoint) diffusion processes can be described using diffusion equations with constant diffusion coefficients in which the partial derivatives with respect to time and coordinate are replaced by fractional-order derivatives. In this case, the subdiffusion process is described by introducing the fractional derivative of order $\alpha$ $(0 < \alpha < 1)$ instead of the partial derivative of the first order in time whereas the superdiffusion process is described by introducing the fractional derivative of order $\beta$ $(0 < \beta < 2)$ instead of the partial derivative of the second order with respect to coordinate [1-4, 11-12]. The parameter $H_1$ is varied in the ranges $0 < H_1 < 0.5$ for subdiffusion and $0.5 < H_1 < 1$ for superdiffusion. It should be noted that if the process under study is more complicated, such as the one in which the diffusion coefficient depends on coordinate, the value of the Hurst constant can be higher than unity.

Presently, there is a substantial amount of data relating various random fluctuations to Eq. (1). Probably the most well-known example is the diffusion of particles in a turbulent flow, i.e., Richardson diffusion [13], which is described by Eq. (1) with $H_1 = 1.5$. There are examples of the occurrence of anomalous diffusion in the charge transfer in semiconductors, dynamics of biological and polymer systems, mass transfer in porous glass, and quantum optics [2, 3, 7-10].

It should be noted that Eq. (1) describes random fluctuations in a virtually unlimited medium and cannot be used to interpolate the mean-square-displacement function for a stochastic process in which the dynamic variable under study approaches a steady or quasi-steady state after some time interval (lag). On the other hand, there are many complex processes, such as the ones analyzed in this paper and Ref. [14], for which the mean square displacement of system states $V(t)$ stops growing after some time. This implies that the dynamics of these systems reaches a steady or quasi-steady state, and their mean square displacements cannot be adequately described using interpolation (1). Therefore, a more general interpolation expression accounting for both the initial exponential growth of the mean square displacement and its behavior near a steady state is needed to describe the dynamics of these complex processes and estimate its parameters.

One of such processes, the dynamics of human magnetoencephalographic (MEG) signals, was recently studied in Ref. [14] using Flicker-Noise Spectroscopy (FNS) [15-19], a phenomenological approach to the analysis of time series. The dynamic characteristics of neuromagnetic responses were determined by analyzing the MEG signals recorded as the response of a group of control human subjects and a patient with photosensitive epilepsy to equiluminant flickering stimuli of different color combinations. It was shown that the mean square displacement of the signals in the control subjects can be described using Eq. (1) at small time lags (displacements), but at larger time lags it becomes proportional to the variance of the source signals. It was therefore suggested that anomalous diffusion can take place in the MEG signals in healthy subjects, implying that the healthy organisms can suppress the perturbations brought about by the flickering stimuli and reorganize themselves. The phenomenological "random walk" ("jump") difference-moment interpolation (Eq. (6) in Ref. [14] or Eq. (6) in this paper) adequately described the stochastic part of the structural function for the source signals. The conclusion about the presence of anomalous diffusion was drawn by relating the "random walk" interpolation at $H_1 = 0.5$ to the mean-square displacement determined from the one-dimensional diffusion equation with symmetry boundary conditions and the Dirac delta function as initial condition. In this case, a good agreement between the interpolation and diffusion model was observed only in the asymptotic cases (at small time intervals and at steady state). The deviation in the intermediate range was substantial, which makes it impossible to use the diffusion equation with symmetry boundary conditions for modeling the stochastic component of MEG and other natural signals.

In this study, we show that the phenomenological random walk interpolation at $H_1 = 0.5$ is practically equivalent to the one-dimensional diffusion equation with integrodifferential boundary conditions incorporating the effects of nonstationarity and the finite residence times of the diffusion system in boundary "adstates". This makes it possible to use the fractional-derivative formulation of the diffusion equation to model the stochastic component of some natural signals with anomalous diffusion. It is also demonstrated that the random walk interpolation adequately describes the stochastic part of the difference moment for four other applications: blinking fluorescence of quantum dots, X-ray emission from accreting objects, fluid velocity in Rayleigh-Bénard convection, and geoelectrical signal for a seismic area. The above implies that the random walk interpolation can be used as a general analytical expression for identifying the anomalous diffusion in complex signals and estimating its parameters, with Eq. (1) being its particular case at small time lags.

**II. THEORY**

In FNS parameterization, the original signal is separated into three components: system-specific "resonances" and their interferential contributions at lower frequencies, stochastic "random walk" ("jump") component at larger frequencies, and stochastic "spike" (inertial) component in the highest frequency range [16]. The parameterization is based on the analysis of the information contained in the autocorrelation function

$$\psi(\tau) = \langle V(t)V(t+\tau) \rangle, \tag{2}$$

where $\tau$ is the time lag parameter. The angular brackets in relation (2) stand for the averaging over time interval $T$:

$$\langle (...) \rangle = \frac{1}{T} \int_{-T/2}^{T/2} (...) dt. \tag{3}$$

To extract the information contained in $\psi(\tau)$ ($\langle V(t) \rangle = 0$ is assumed), the following transforms, or "projections", of this function are analyzed:

cosine transforms (power spectrum estimates) $S(f)$, where $f$ is the frequency

$$S(f) = \int_{-T/2}^{T/2} \langle V(t)V(t+t_1) \rangle \cos(2\pi f t_1) dt_1 \tag{4}$$

and its difference moments (Kolmogorov transient structural functions) of the second order $\Phi^{(2)}(\tau)$

$$\Phi^{(2)}(\tau) = \langle [V(t) - V(t+\tau)]^2 \rangle. \tag{5}$$

The basic idea of FNS parameterization is that the correlation links present in sequences of different irregularities, such as spikes (Dirac delta function), "jumps" (Heaviside function), and discontinuities in derivatives of different orders, on all levels of the spatiotemporal hierarchy of the system under study can be treated as the main carriers of information about the stochastic processes in the signal. It is further assumed that according to the self-organized criticality paradigm [20], the stochastic dynamics of real processes is associated with intermittency, consecutive alternation of rapid changes in the values of dynamic variables on small time intervals with small variations of the values on longer time intervals. Such intermittency occurs on every hierarchical level of the system evolution. It is also assumed that the individual features of the evolution of complex systems are mostly contained in low-frequency (regular) components of the signals; i.e., internal and external resonances and their interferential contributions. The regular components are found in the background of stochastic ("noise") high-frequency components associated with spike and jump irregularities of dynamic variable $V(t)$. The sequences of spike and jump irregularities reflect the intermittent nature of stochastic evolution with spikes corresponding to rapid fluctuations between laminar phases and jumps corresponding to "random walks", random stepwise changes in dynamic variable $V(t)$ in the region of laminar phases (Fig. A.1). As demonstrated in Appendix A, the information contents of $S(f)$ and $\Phi^{(2)}(\tau)$ for this intermittent signal are different, and the parameters for both functions are needed to solve the parameterization problem [16, 17, 21]. Interpolation expressions for the stochastic components of $\Phi^{(2)}(\tau)$ and $S(f)$, $\Phi_s^{(2)}(\tau)$ and $S_s(f)$, respectively, that take the intermittent character of the signals under study into account were derived using the theory of generalized functions in Ref. [17]. It was shown that the structural functions $\Phi^{(2)}(\tau)$ are formed only by jump irregularities, and that the functions $S(f)$, which characterize the "energy side" of the process, are formed by both types of irregularities, spikes and jumps.

The complete parameterization procedure using discrete expressions is listed in Appendix B. A detailed discussion of the algorithm is presented elsewhere [16, 21].

The random walk interpolation for the stochastic part of the difference moment derived using the theory of generalized functions in Ref. [17] for the case characterized by the same dynamic-correlation parameters on every evolution hierarchy level and only one characteristic scale in the sequences of spikes and jumps is written as:

$$\Phi_s^{(2)}(\tau) \approx 2\sigma^2 \cdot \left[1 - \Gamma^{-1}(H_1) \cdot \Gamma(H_1, \tau/T_1)\right]^2, \tag{6}$$

$$\Gamma(s,x) = \int_x^\infty \exp(-t) \cdot t^{s-1} dt, \quad \Gamma(s) = \Gamma(s,0) \quad.$$

Here, the subscript $s$ stands for the stochastic part of the signal; $\Gamma(s)$ and $\Gamma(s,x)$ are the complete and incomplete gamma functions ($x \geq 0$ and $s > 0$), respectively; $\sigma$ is the standard deviation of the measured dynamic variable with dimension $[V]$; $H_1$ is the Hurst constant, which describes the rate at which the dynamic variable "forgets" its values on the time intervals that are less than the correlation time $T_1$. In this case, $T_1$ may be interpreted as the correlation time for the jumps in stochastically varying time series $V(t)$. Expression (6) was derived using the probability density function for jump irregularities presented in Eq. (23) in Ref. [17], which is characterized by a maximum and an exponential decline at large argument values. This choice of the probability density function made it possible to describe the loss of correlation links in a sequence of jump irregularities using two phenomenological parameters $H_1$ and $T_1$. The third phenomenological parameter $\sigma$ in this case characterizes the "intensity" of the irregularities.

For asymptotic cases, we come to the expressions [17]:

$$\Phi_s^{(2)}(\tau) = 2\Gamma^{-2}(1+H_1) \cdot \sigma^2 \left(\frac{\tau}{T_1}\right)^{2H_1}, \quad \text{if } \frac{\tau}{T_1} \ll 1; \tag{7}$$

$$\Phi_s^{(2)}(\tau) = 2\sigma^2 \left[1 - \Gamma^{-1}(H_1) \cdot \left(\frac{\tau}{T_1}\right)^{H_1-1} \exp\left(-\frac{\tau}{T_1}\right)\right]^2, \quad \text{if } \frac{\tau}{T_1} \gg 1. \tag{8}$$

Although expressions (6)-(8) were derived for a statistically stationary signal in which the phenomenological parameters are the same on each level of the system evolution hierarchy, they can also be applied to the analysis of real signals, which are generally nonstationary. To do it, the real signals at specific averaging intervals and sampling frequencies should be regarded as quasi-stationary with a specific variance and other phenomenological parameters. In this case, the values of the phenomenological parameters may vary at different quasi-stationary intervals (see Example I.A).

FNS is a phenomenological approach. The parameters introduced in FNS have a certain physical meaning and are determined by comparing the results calculated by FNS interpolations with the curves calculated from the experimental values of time series $V(t)$. Consider a stationary process in which the autocorrelator $\psi(\tau) = \langle V(t)V(t+\tau) \rangle$ depends only on the difference in the arguments of dynamic variables and the ergodicity condition is met. For this stationary process, the FNS procedure of averaging over time, (3), is equivalent to the averaging procedure using the probability density function $W(V, t)$ for finding the values of the dynamic variable in the interval from $V$ to $V + dV$ at time $t$. In this case, expression (6) may be regarded as the generalized expression for the mean square displacement in the random walk processes described by Fickian equation or the equations of anomalous diffusion.

In Ref. [14], it was shown that expression (6) in the case of Fickian diffusion ($H_1 = 0.5$) has the same asymptotic values as the mean-square displacement determined from the one-dimensional diffusion equation with symmetry boundary conditions and the Dirac delta function as the initial condition. At the same time, the deviation in the intermediate time range is as high as 20%. This is attributed to the limitations of symmetry boundary conditions with regard to real systems. These limitations can be eliminated by using more general boundary conditions.

Assume that the behavior of the probability density $W(V, \tau)$ (we use $\tau$ rather than $t$ as the process is stationary, with the start time $\tau = 0$) for the random variable $V$ on the segment $[-L, +L]$ over time $\tau$ can be described by the diffusion equation:

$$\frac{\partial W}{\partial \tau} = D \frac{\partial^2 W}{\partial V^2}. \qquad (9)$$

In this case, the boundaries of interval $[-L, L]$ determine the standard deviation of the dynamic variable from the mean value $V = 0$. The use of symmetry boundary conditions in Ref. [14] implies that the values of the dynamic variable are locked in the interval. To consider the real dynamics of changes in $V(\tau)$, it is necessary to allow the values of $V(\tau)$ to go outside the interval and stay there for some finite time prior to coming back. Such "retardation" effects can be taken into account by using more general boundary conditions [22] in which additional "delay" processes are incorporated by introducing "adstates" at the boundaries $+L$ and $-L$, where the system may stay a finite time. Introducing the probability densities $w_{\pm L}(\tau)$ for finding the system in such boundary "adstates", the boundary conditions can be written as:

At $V = -L$:

$$D\frac{\partial W}{\partial V} = \chi W(-L,\tau) - \lambda w_{-L}(\tau), \tag{10}$$

$$\frac{dw_{-L}(\tau)}{dt} = \chi W(-L,\tau) - \lambda w_{-L}(\tau). \tag{11}$$

At $V = +L$:

$$-D\frac{\partial W}{\partial V} = \chi W(+L,\tau) - \lambda w_{+L}(\tau), \tag{12}$$

$$\frac{dw_{+L}(t)}{dt} = \chi W(+L,\tau) - \lambda w_{+L}(\tau). \tag{13}$$

Here, $\chi$ and $\lambda$ are the rate constants for direct and reverse transitions of the system from a boundary "diffusion" state to an "adstate", respectively. Assume that their values are the same for both boundaries.

After solving Eqs. (11) and (13) for probability densities $w_{-L}(\tau)$ and $w_{+L}(\tau)$ and substituting the solutions into Eqs. (10) and (12), the boundary conditions take the form:

$$D\frac{\partial W}{\partial V} = \chi W(-L,\tau) - \lambda\chi\exp(-\lambda\tau)\int_0^\tau W(-L,\xi)\exp(\lambda\xi)d\xi \quad \text{at} \quad V = -L, \tag{14}$$

$$-D\frac{\partial W}{\partial V} = \chi W(L,\tau) - \lambda\chi\exp(-\lambda\tau)\int_0^\tau W(L,\xi)\exp(\lambda\xi)d\xi \quad \text{at} \quad V = +L. \tag{15}$$

When $\chi = 0$ and $\lambda = 0$, boundary conditions (14)-(15) transform to the traditional symmetry boundary conditions:

$$\frac{\partial W}{\partial V} = 0 \text{ at } V = -L \text{ and } V = +L. \tag{16}$$

The following initial condition will be used:

$$W(V,0) = \delta(V). \tag{17}$$

Other approaches to generalizing the mathematical model given by diffusion equation (9) and conditions (16)-(17) are possible. These include the introduction of a variable diffusion coefficient, the addition of the second-order partial derivative of the probability density $W(V, \tau)$ with respect to time to the left side of Eq. (9) – "telegraph equation", and others. However, the generalization of symmetry boundary conditions is the only approach that is physically substantiated and simultaneously meets the assumptions used in deriving Eq. (6).

We will use an iterative method for solving the integrodifferential problem given by Eqs. (9), (14)-(15), (17). At the initial step, we will approximate $W(L,\xi)$ and $W(-L,\xi)$ in the boundary conditions with the expression

$$W(V,\tau) = \frac{1}{2L}\left[1 + 2\sum_{k=1}^{\infty} \exp\left(-\frac{\pi^2 k^2 D\tau}{L^2}\right) \cos\frac{\pi k V}{L}\right], \tag{18}$$

which is the solution for the case of symmetry boundary conditions [14].

At the boundaries, expression (18) takes the following form:

$$W(-L,\tau) = W(L,\tau) = \frac{1}{2L}\left[1 + 2\sum_{k=1}^{\infty} (-1)^k \exp\left(-\frac{\pi^2 k^2 D\tau}{L^2}\right)\right]. \tag{19}$$

Considering the facts that the first term in the sum approximates well the total sum for most of the time interval because $\exp(-k^2)$ dramatically decreases and that the value of $W$ is close to zero at the initial time interval, where $1 - 2\exp\left(-\frac{\pi^2 D\tau}{L^2}\right) < 0$, we can use the approximation:

$$W(-L,\tau) = W(L,\tau) \approx \frac{1}{2L} \times \Theta\left[1 - 2\exp\left(-\frac{\pi^2 D\tau}{L^2}\right)\right] \times \left\{1 - 2\exp\left(-\frac{\pi^2 D\tau}{L^2}\right)\right\}. \qquad (20)$$

Here, $\Theta[x]$ is the Heaviside function, which is equal to "0" for negative and "1" for nonnegative argument values.

Comparison of analytical expression (20) and a numerical solution to the corresponding problem given by (9), (16), and (17) shows that the approximation is in good agreement with the numerical solution (Fig. 1). The numerical solution (dashed curved in Fig. 1) was obtained using a built-in MATLAB procedure, *pdepe*, for solving initial-boundary problems with one-dimensional parabolic-elliptic partial differential equations. The *pdepe* procedure is based on the discretization of partial differential equations in space using second order approximations and subsequent time integration of the resulting ordinary differential equations. The space discretization step was chosen to match the initial condition based on the Dirac delta function.

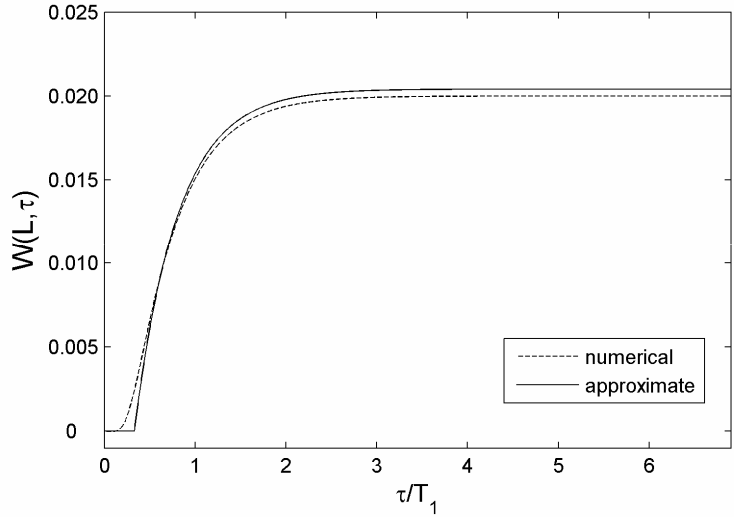

FIG. 1. Comparison of Eq. (20) with the numerical solution for symmetry boundary conditions (16).

By evaluating the integrals in boundary conditions (14) and (15) with $W(L,\xi)$ and $W(-L,\tau)$ presented in Eq. (20), we obtain

$$D\frac{\partial W}{\partial V} = \chi W(-L,\tau) - \frac{\lambda\chi \exp(-\lambda\tau)}{2L} \times \Theta\left[1 - 2\exp\left(-\frac{\pi^2 D\tau}{L^2}\right)\right] \times \Omega(\tau), \qquad (21)$$

$$-D\frac{\partial W}{\partial V} = \chi W(L,\tau) - \frac{\lambda\chi \exp(-\lambda\tau)}{2L} \times \Theta\left[1 - 2\exp\left(-\frac{\pi^2 D\tau}{L^2}\right)\right] \times \Omega(\tau), \qquad (22)$$

where

$$\Omega(\tau) = \left[\left\{\frac{\exp(\lambda\tau) - \exp\left(\frac{\ln(2)\lambda L^2}{\pi^2 D}\right)}{\lambda}\right\} - 2\frac{\exp\left(\left\{\lambda - \frac{\pi^2 D}{L^2}\right\}\tau\right) - \exp\left(\left\{\lambda - \frac{\pi^2 D}{L^2}\right\}\frac{\ln(2) L^2}{\pi^2 D}\right)}{\lambda - \frac{\pi^2 D}{L^2}}\right].$$

To solve the problem given by (9), (14)-(15), (17), we first numerically solved the problem given by (9), (17), (21)-(22) using the *pdepe* function in MATLAB, and then substituted the solution for $W(L,\tau)$ and $W(-L,\tau)$ into boundary conditions (14) and (15). The resulting problem was solved numerically. The procedure was iterated until the required accuracy for $W(L,\tau)$ was achieved.

To compare interpolation (6) with the mean-square displacement calculated using the solution to the problem given by (9), (14)-(15), (17), the same normalized dimensionless functions as in Ref. [14], which were obtained by matching the asymptotic expressions at $\tau/T_1 \ll 1$ and $\tau/T_1 \gg 1$, respectively, were used:

$$\phi_1(\tau) = \frac{3}{L^2}\langle V^2 \rangle_{pdf}, \quad \phi_2(\tau) = \frac{1}{2\sigma^2}\Phi_s^{(2)}(\tau); \qquad (23)$$

where

$$\langle V^2 \rangle_{pdf} = \frac{\int\limits_{-L}^{+L} V^2 W(V,\tau)dV}{\int\limits_{-L}^{+L} W(V,\tau)dV} \qquad (24)$$

is the mean-square-displacement function.

As can be seen in Fig. 2, the normalized mean-square-displacement curve calculated for integrodifferential boundary conditions (14)-(15) runs much closer to interpolation formula (6) than does the curve for symmetry boundary conditions (16). The relative error for intermediate values of $\tau$ in this case does not exceed 5%. In other words, interpolation (6) is practically equivalent to the mean-square-displacement function calculated by solving the problem given by (9), (14)-(15), (17); that is, the diffusion model with integrodifferential boundary conditions.

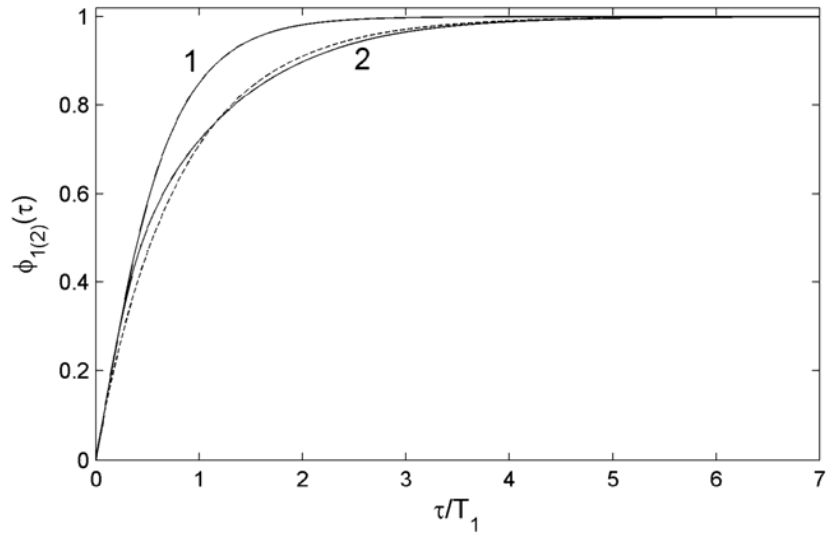

FIG. 2. Normalized mean-square-displacement functions $\phi_1(\tau)$ [curve 1, symmetry boundary conditions (16); curve 2, integrodifferential boundary conditions (21)-(22) at $\chi = 0.4$, $\lambda = 0.04$] and $\phi_2(\tau)$ [dashed line, interpolation formula (6)].

It follows from Fig. 3, which illustrates the variations of $W$ at the boundaries and the mean-square-displacement function for the initial, first, and last iterations, that the deviation of the boundary values of $W(V, \tau)$ for the symmetry boundary conditions (initial iteration) from the boundary values for the integrodifferential conditions is so large that it takes several dozen iterations for the numerical solution to the problem given by (9), (14)-(15), (17) to reach the required accuracy.

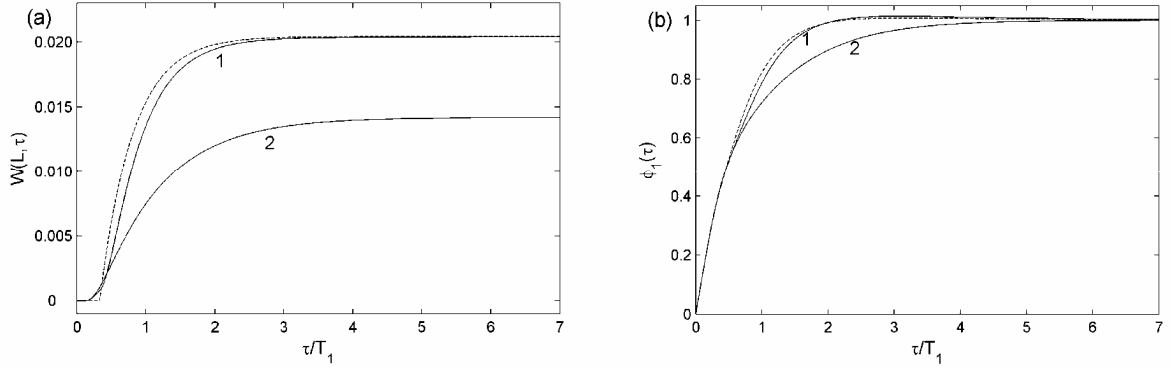

FIG. 3. (a) Probability density functions $W(L,\tau)$ and (b) normalized mean-square-displacement functions $\phi_1(\tau)$ for the initial (dashed), first (curve 1), and last, $50^{th}$ (curve 2), iterations.

The analysis of several natural signals with $H_1 = 0.5$, such as Fig. 8 in Ref. [14], demonstrated that phenomenological expression (6), which is practically equivalent to the diffusion problem with integrodifferential boundary conditions, describes the stochastic component of experimental difference moments much better than the model expression derived from the diffusion problem with symmetry boundary conditions (16).

The above facts imply that the mathematical models for the stochastic processes in real complex systems with anomalous diffusion can be developed on the basis of the equations of anomalous diffusion with integrodifferential boundary conditions (14)-(15).

It can be easily shown that the asymptotic expressions for the problem with integrodifferential boundary conditions are the same as for the problem with symmetry boundary conditions because the value of $W$ at the boundaries is constant at both the initial time interval and steady state, which can be seen in Fig. 3(a). The asymptotic expressions are written as [14]:

$$\langle V^2 \rangle_{pdf} \to 2D\tau \qquad \text{when } \tau \ll \frac{L^2}{\pi^2 D}, \qquad (25)$$

$$\left\langle V^2 \right\rangle_{pdf} \to \frac{L^2}{3} \qquad \text{when } \tau \gg \frac{L^2}{\pi^2 D}. \tag{26}$$

The relations between the parameters of the diffusion problem and the phenomenological FNS parameters in interpolation (6) can be found by comparing asymptotic expressions (25) and (26) for small and large values of $\tau$ with the corresponding expressions (7) and (8) written for $H_1 = 0.5$:

$$D = \frac{4}{\pi} \cdot \frac{\sigma^2}{T_1}; \quad L^2 = 6\sigma^2. \tag{25}$$

It is difficult to compare the functions $<V^2>_{pdf}$ and $\Phi^{(2)}(\tau)$ for anomalous diffusion ($H_1 \neq 0.5$) because the corresponding equations for the probability density $W(V, \tau)$ of random variable $V$ varying on the segment $[-L, +L]$ over time $\tau$ are very complicated and can be solved only by numerical methods [3, 4, 9]. At the same time, the desired relation can be found if we compare expressions (7) and (1), which correspond to the case of anomalous diffusion at small values of $\tau$, by choosing $T_1$ as the characteristic time $t_0$, and assume that in this case the FNS parameter $\sigma$ corresponds to some model parameter $L_a$ determining the region in the range of values of the dynamic variable where the states of the system can be localized:

$$D = \frac{1}{\Gamma^2(1+H_1)} \cdot \frac{\sigma^2}{T_1}; \quad L_a^2 = b\sigma^2, \tag{26}$$

Here, $b$ is a dimensionless parameter.

To extract the "random walk" component from the time series analyzed above, we used the FNS parameterization algorithm listed in Appendix B and described in detail in Refs. [16, 21]. In addition to the random walk parameters $\sigma$, $H_1$, and $T_1$, the values of three other FNS parameters, which are related to the stochastic "spike" component, are given in the figure captions: $S_c(0)$, the low-frequency limit of the stochastic power spectrum estimate; $T_0$, the correlation time for spikes; and $n$, the degree of correlation loss in the sequence of spikes on time interval $T_0$. More detailed description of these parameters is presented elsewhere [16, 21].

## III. EXAMPLES

**A. Anomalous diffusion in blinking fluorescence of quantum dots**

The recent progress in nanotechnologies gave rise to several new problems related to the standardization and stabilization of the functional activity of quantum-sized objects. Particular examples are "quantum dots" (QD), which are regarded as the functional elements of the quantum computers of the future, and photoactive elements, which are added or formed in inorganic or organic matrices during the production of optical materials (photochromes, luminophores).

We will analyze the fluorescence signal generated by a CdSe quantum dot overcoated with a thin layer of ZnS (so-called core-shell QD) [23, 24]. The experimental data were kindly provided by Professor Masaro Kuno (University of Notre Dame). The radius of the core CdSe particle was 2.7 nm. There were 3 "monolayers" of ZnS surrounding this particle. The laser excitation wavelength was 488 nm and the incident intensity was 600 W/cm$^2$. The absorption cross section of such dots was approximately $10^{-15}$ cm$^2$. The quantum yield of these dots at the ensemble level should be of the order of ~30%. The signal was recorded during $T_{tot}$ = 1 hour at a sampling frequency $f_d$ = 100 Hz. So, the time series contains 360,000 values in the time interval of $1 - 360000\, f_d^{-1}$. During this time interval the fluorescence intensity significantly dropped. To demonstrate the nonstationarity of the signal, which manifests itself in the dependence of the FNS parameters on the averaging window selected within the interval $T_{tot}$, in addition to the source signal we also analyzed its three slices with 36,000 values each, corresponding to the initial, intermediate, and final stages.

Figures 4(a)-7(a) show the source signal and its slices at intervals I ($1 - 36000\, f_d^{-1}$), II ($162001 - 198000\, f_d$) and III ($324001 - 360000\, f_d$), respectively. The graphs of the relations used in the FNS analysis are presented in Figs. 4(b)-(d) – 7(b)-(d), with the FNS parameters being given in the figure captions. To reduce the fitting error in describing the experimental difference moment, the zeroth frequency point was excluded from the regular difference moment component, i.e., $q_{\min}=1$ was used in the parameterization algorithm listed in Appendix B. It can be seen that the stochastic part $\Phi_c^{(2)}(\tau)$ of experimental difference moments for all four signals is adequately approximated by anomalous-diffusion interpolation (6), which in this case corresponds to "subdiffusion". The differences in the values of FNS parameters for the slices at intervals I, II, and III imply that the signal is nonstationary. Fifty fluorescence signals produced by other single QDs were also analyzed (not listed here). For some of the signals, anomalous-diffusion interpolation (6) could not adequately describe the stochastic part of the experimental difference moment. This can be attributed to the fact that the signals under study were recorded for a highly nonstationary process (at a relaxation stage) or were characterized by two or more scales, which are two general situations when interpolation formula (6) may be inadequate. In the latter case, one can use interpolation expressions with higher numbers of parameters (see Ref. [25] for details). In the case of complex relaxation dynamics, one can introduce phenomenological exponential interpolation expressions with parameters nonlinearly changing with time and then use standard FNS parameterization (see Appendix B) to study the dynamics of the variable with reference to the exponential interpolation.

The above analysis demonstrates that the stochastic dynamics in the nanoscale complex system under study is governed by anomalous diffusion.

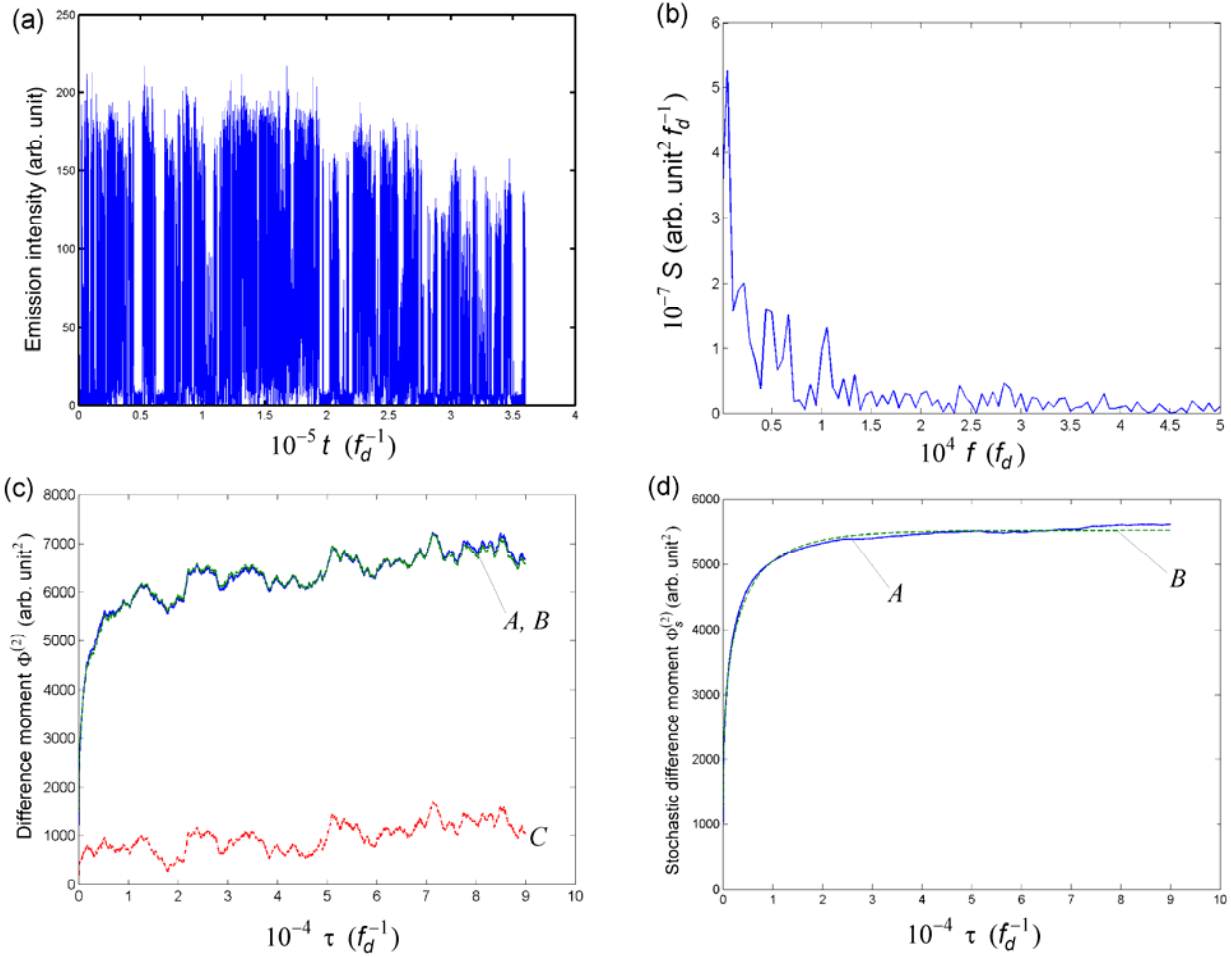

FIG. 4. (Color online) Analysis of the fluorescence signal generated by a CdSe quantum dot overcoated with a thin layer of ZnS (so-called core-shell QD) on the time interval from 1 to 360000 $f_d^{-1}$ ($\sigma$ = 52.6 arb. units; $H_1$ = 0.12; $T_1$ = 13269 $f_d^{-1}$ ≈ 133 s, $D$ ≈ 23.4 (arb. units)$^2$/s, $S_s(0)$ = 3.10×10$^7$ (arb. units)$^2 f^{-1}{}_d$; $n$ = 1.25; $T_0$ = 7114 $f_d^{-1}$ ≈ 71.1 s): (a), source signal; (b), power spectrum $S(f)$ presented in Eq. (A.2) in the low-frequency range; (c), $A$ [solid blue line] – experimental $\Phi^{(2)}(\tau)$ presented in Eq. (A.8), $B$ [dashed green line] – general interpolation for $\Phi^{(2)}(\tau)$ given by the sum of expressions in Eqs. (A.7) and (A.10), $C$ [dot-dashed red line] – $\Phi_r^{(2)}(\tau)$ presented in Eq. (A.7); (d), $A$ [solid blue line] – $\Phi_s^{(2)}(\tau)$ presented in Eq. (A.9), $B$ [dashed green line] – stochastic interpolation $\Phi_s^{(2)}(\tau)$ presented in Eq. (6).

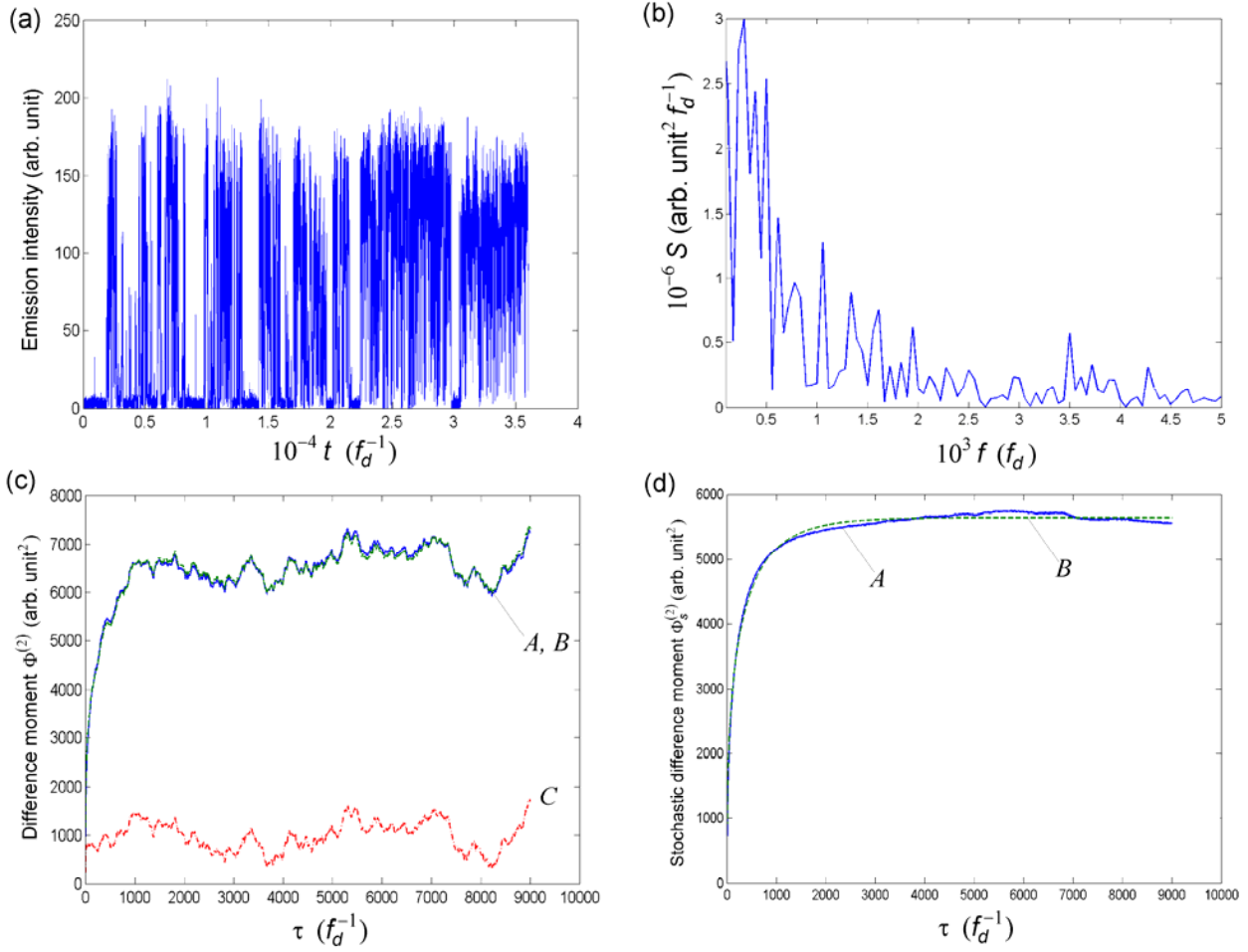

FIG. 5. (Color online) Analysis of the signal in Fig. 4 on the time interval from 1 to 36000 $f_d^{-1}$ ($\sigma$ = 53.1 arb. units; $H_1$ = 0.18; $T_1 = 968\, f_d^{-1} \approx 9.7$ s, $D \approx 341.6$ (arb. units)$^2$/s, $S_s(0) = 3.27 \times 10^6$ (arb. units)$^2 f^{-1}{}_d$; $n$ = 1.40; $T_0 = 513\, f_d^{-1} \approx 5.13$ s; $q_{min}$ =1 - see Appendix B). The subfigure captions are the same as in Fig. 4.

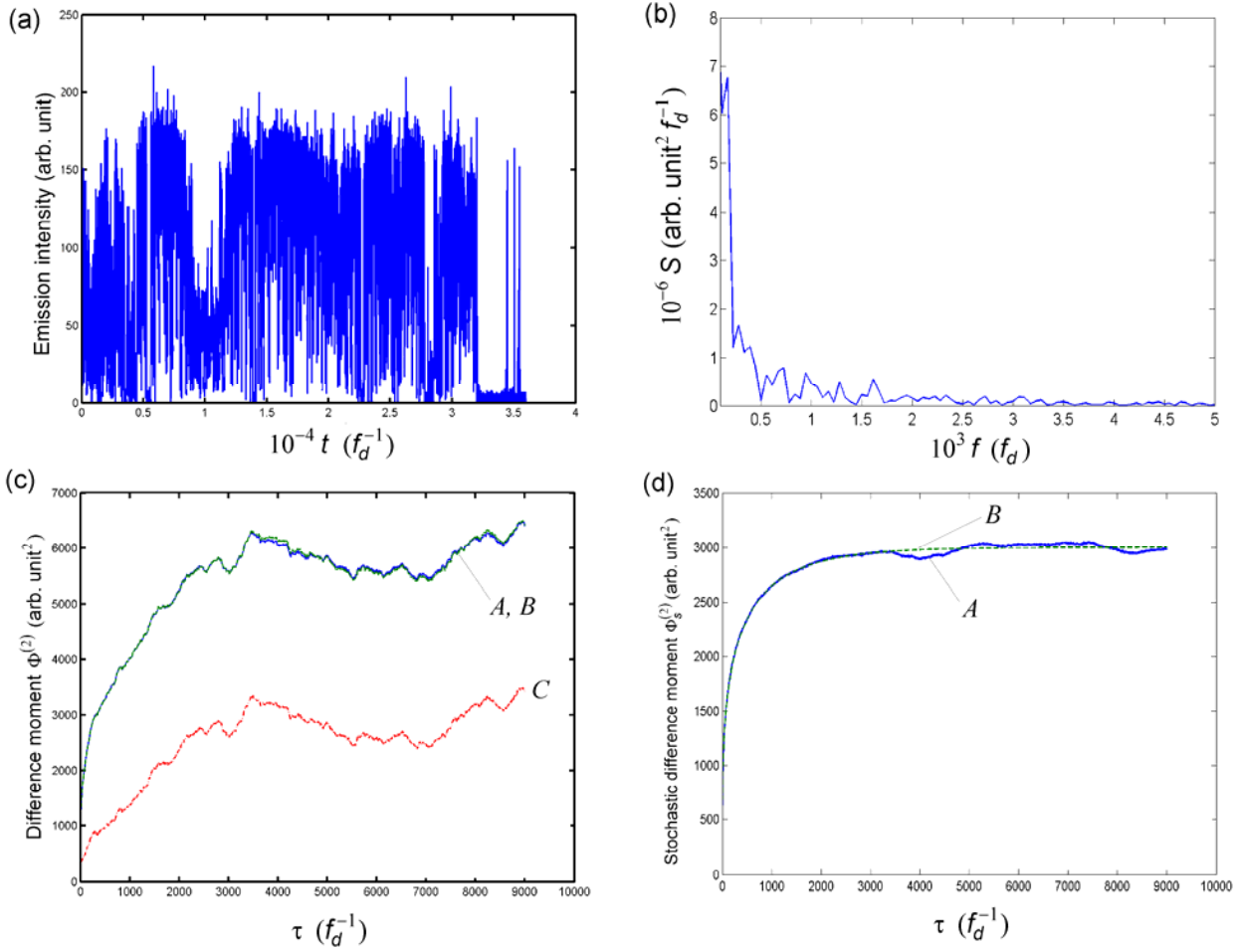

FIG. 6. (Color online) Analysis of the signal in Fig. 4 on the time interval from 162000 to 198000 $f_d^{-1}$ ($\sigma$ = 38.7 arb. units; $H_1$ = 0.13; $T_1$ = 1585.7 $f_d^{-1}$ ≈ 15.86 s, $D$ ≈ 107.2 (arb. units)$^2$/s, $S_s(0)$ = 9.35×10$^6$ (arb. units)$^2 f_d^{-1}$; $n$ = 1.17; $T_0$ = 4028.3 $f_d^{-1}$ ≈ 40.3 s; $q_{min}$ = 1 - see Appendix B). The subfigure captions are the same as in Fig. 4.

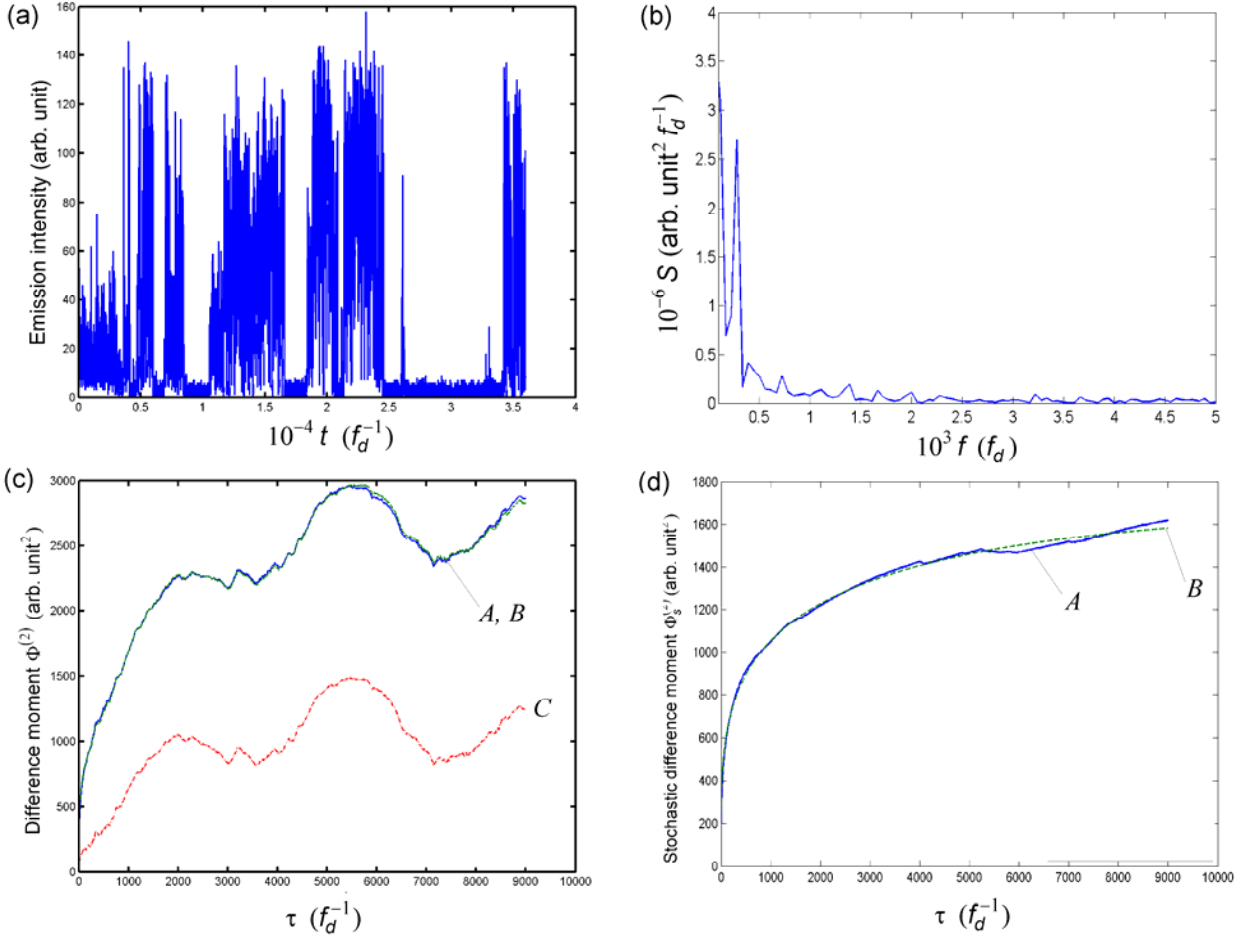

FIG. 7. (Color online) Analysis of the signal in Fig. 4 on the time interval from 324000 to 360000 $f_d^{-1}$ ($\sigma$ = 29.2 arb. units; $H_1 = 0.13$; $T_1 = 10077 f_d^{-1} \approx 100.1$ s, $D \approx 9.56$ (arb. units)$^2$/s, $S_s(0) = 3.88 \times 10^6$ (arb. units)$^2 f^{-1}{}_d$; $n = 1.23$; $T_0 = 3547 f_d^{-1} \approx 35.5$ s; $q_{min} = 1$ - see Appendix B). The subfigure captions are the same as in Fig. 4.

### B. Anomalous diffusion in the dynamics of X-ray emission from accreting objects

This section shows that anomalous diffusion, as the process of stochastic state changes of a complex system, can manifest itself in the dynamics of stellar objects. As an example, we will consider the dynamics of X-ray emission from two accreting stellar systems, GRS 1915+105 and Cygnus X-1.

The binary system GRS 1915+105, which is located in the Aquila constellation approximately 40,000 light years away from the Sun, is a star-donor with mass $M_d = 1.2 \pm 0.2\ M_{sun}$, where $M_{sun}$ is the mass of the Sun, that revolves around a spinning compact heavy object, the black hole with mass $M_{bh}$ = 14 ± 4 $M_{sun}$. The orbital period of this system is 33.5 ± 1.5 days. The interest of researchers to this system [26-31] was brought about by the high luminosity of its accretion disk, which is close to the Eddington limit, when the radiation pressure on the accreting matter is comparable to the gravitational attraction to the central object. This leads to unstable modes of matter transfer, which produce powerful X-ray outbursts and relativistic beams of particles (jets). For this reason, GRS 1915+105 is considered as a "microquasar", the stellar analog of active galactic nuclei [26].

The other microquasar, the binary system Cygnus X-1, is a powerful source of X-ray emission, which is located 6,000 light years away from the Sun. The optical component of this system is a blue supergiant variable star with surface temperature around 31,000 K and the mass of 33 ± 9 $M_{sun}$. The lower limit of the mass of the accreting object, a black hole into which the matter flows from the atmosphere of the supergiant resulting in a flat gas disk, is estimated as 16 ± 5 $M_{sun}$.

X-ray emission (impulses with various powers and durations – up to milliseconds) is generated in the inner layers of the gas disk, the temperatures of which are estimated to be of the orders of $10^7$-$10^8$ K [28].

The time series $I(t)$ for the total flow of X-ray emission from these sources (the primary data are available on the Internet [32]) in the period from January 1, 1996 to December 31, 2005 are shown in Figs. 8(a) and 9(a). The average intervals between adjacent measurements were $\Delta t_{grs}$ = 106 minutes and $\Delta t_{cygx}$ = 88 minutes, with the corresponding sampling frequencies $f_d$ of $1.57 \times 10^{-4}$ $Hz$ and $1.89 \times 10^{-4}$ $Hz$; the total numbers of measurements were 49,355 and 59,748 for GRS 1915+105 and Cygnus X-1, respectively. The errors in the measurements were attributed to the variation of the intervals between adjacent measurements: approximately in 10% of the cases the intervals deviated from the average values $\Delta t_{grs}$ and $\Delta t_{cygx}$ by two times. The dynamics of the X-ray emission sources was previously studied in Refs. [27-29]. Based on the analysis of probability density functions and nonlinear dynamics of complex systems, it was suggested that the physical mechanisms of matter transfer are different between the systems and concluded that the self-similarity is limited in the dynamics of transfer [27, 28]. It was shown that the signals produced by Cygnus X-1 during the transfer of matter in the accretion disk could be characterized using a model of anomalous diffusion with Hurst constant $H_1 \approx 0.3$ and time self-similarity on interval $T_s \approx 3$ years. Some of the features of transfer processes in the accretion disk of GRS 1915+105 were similar to the corresponding processes in Cygnus X-1: $H_1 \approx 0.35$, which also pointed to "subdiffusion". However, the time self-similarity was seen only on the intervals of $T_s \approx$ 12-17 days.

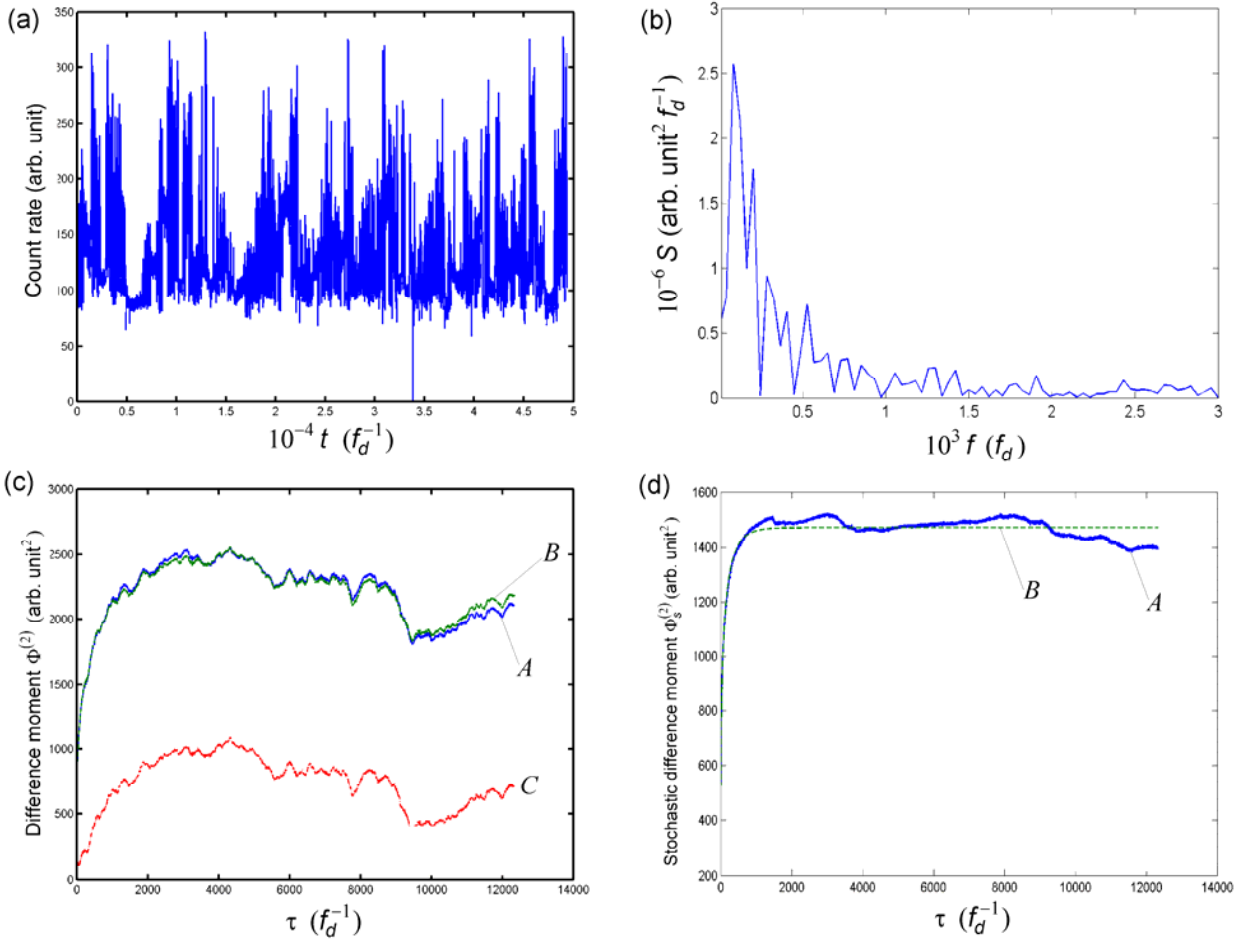

FIG. 8. (Color online) Analysis of the dynamics of X-ray emission from the binary system GRS 1915+105 in the period from January 1, 1996 to December 31, 2005 [32] ($\sigma = 27.1$ arb. units; $H_1 = 0.13$; $T_1 = 396$ $\Delta t_{grs} \approx 29$ days, $D \approx 28.6$ (arb. units)$^2$/day, $S_s(0) = 5.57 \times 10^5$ (arb. units)$^2 f^{-1}{}_d$; $n = 1.13$; $T_0 = 568.5$ $\Delta t_{grs} \approx 42$ days): (a), source signal; (b), power spectrum $S(f)$ presented in Eq. (A.2) in the low-frequency range; (c), $A$ [solid blue line] – experimental $\Phi^{(2)}(\tau)$ presented in Eq. (A.8), $B$ [dashed green line] – general interpolation for $\Phi^{(2)}(\tau)$ given by the sum of expressions in Eqs. (A.7) and (A.10), $C$ [dot-dashed red line] – $\Phi_r^{(2)}(\tau)$ presented in Eq. (A.7); (d), $A$ [solid blue line] – $\Phi_s^{(2)}(\tau)$ presented in Eq. (A.9), $B$ [dashed green line] – stochastic interpolation $\Phi_s^{(2)}(\tau)$ presented in Eq. (6).

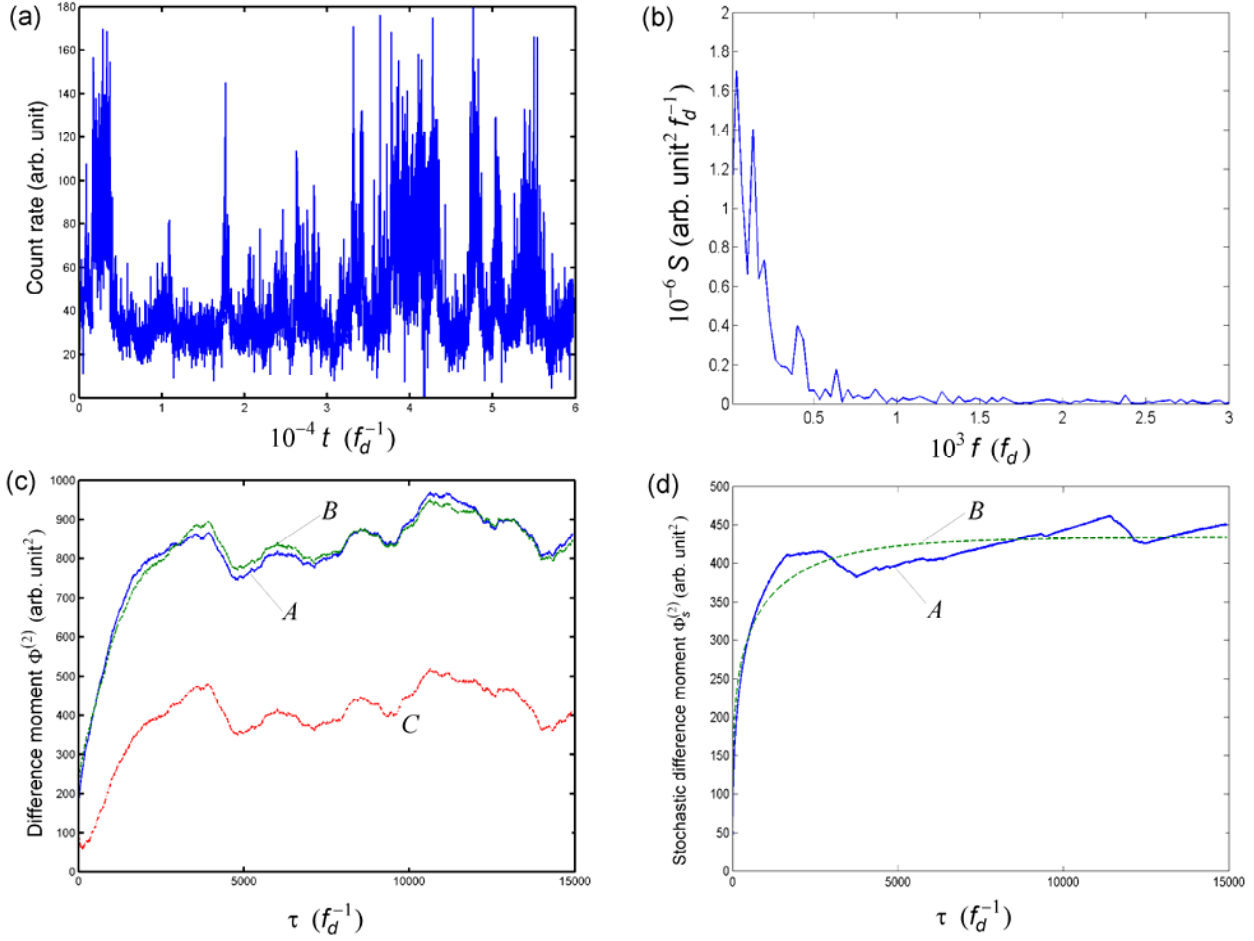

FIG. 9. (Color online) Analysis of the dynamics of X-ray emission from binary system Cygnus X-1 in the period from January 1, 1996 to December 31, 2005 [32] ($\sigma = 14.7$ arb. units; $H_1 = 0.11$; $T_1 = 3444$ $\Delta t_{cygx} \approx 210$ days, $D \approx 1.15$ (arb. units)$^2$/day; $S_s(0) = 8.8 \times 10^5$ (arb. units)$^2 f_d^{-1}$; $n = 1.29$; $T_0 = 2123$ $\Delta t_{cygx} \approx 130$ days): (a), source signal; (b), power spectrum $S(f)$ presented in Eq. (A.2) in the low-frequency range; (c), $A$ [solid blue line] – experimental $\Phi^{(2)}(\tau)$ presented in Eq. (A.8), $B$ [dashed green line] – general interpolation for $\Phi^{(2)}(\tau)$ given by the sum of expressions in Eqs. (A.7) and (A.10), $C$ [dot-dashed red line] – $\Phi_r^{(2)}(\tau)$ presented in Eq. (A.7); (d), $A$ [solid blue line] – $\Phi_s^{(2)}(\tau)$ presented in Eq. (A.9), $B$ [dashed green line] – stochastic interpolation $\Phi_s^{(2)}(\tau)$ presented in Eq. (6).

The results of FNS analysis of the signals shown in Figs. 8(a) and 9(a) are presented in Figs. 8(b)-(d) and 9(b)-(d), with the FNS parameters being given in the figure captions. Though the actual values of $H_1$, 0.13 for GRS 1915+105 and 0.11 for Cygnus X-1, were different from the values given in Refs. [27, 28], 0.35 and 0.3, respectively, the FNS parameters pointed to the same mode of "subdiffusion". It should be noted that the values of parameters $H_1$, $\sigma$, and $T_1$, calculated using the method of non-linear least squares, were chosen by providing the best agreement between the experimental and calculated curves $\Phi^{(2)}(\tau)$ in the entire interval of $\tau$ under study. At the same time, the Hurst constant $H_1$ in Refs. [27, 28] was calculated based on the agreement with experimental data in the limit of small values of $\tau$. More significant differences were seen for the values of $T_1$: 29 days for GRS 1915+105 and 210 days for Cygnus X-1. The differences between the values of $T_1$ and $T_s$ should be mostly attributed to the fact that the methods of analysis used in Refs. [27, 28] do not include the separation of regular components from the signal before performing the stochastic parameterization. This factor was less important in calculating $H_1$ because only the small values of $\tau$ (high-frequency range) were used, where the effect of regular components is minimal. In calculating $T_s$, the low-frequency regular components play an important role and thus should be removed before performing the stochastic parameterization.

The above analysis demonstrates that the stochastic changes in the states of microquasars (that is, astrophysical objects) can be described in terms of anomalous diffusion.

**C. Anomalous diffusion in Rayleigh-Bénard convection**

Rayleigh-Bénard convection, which occurs in the fluid trapped between two horizontal plates when the bottom plate is warmer than the top one, is considered as one of the model problems for studying thermal convection for a broad spectrum of physical phenomena encountered in the Earth's mantle, atmosphere, and oceans [33, 34]. Rayleigh-Bénard convection is characterized by the Rayleigh number $\text{Ra} = g\alpha\Delta T_{rb}H_{rb}/(\kappa\nu)$, Prandtl number $\text{Pr} = \nu/\kappa$, and aspect ratio $\gamma = W_{rb}/H_{rb}$. Here, $g$ is the gravitational acceleration, $\Delta T_{rb}$ is the temperature difference between the plates, $H_{rb}$ is the distance between the plates, $W_{rb}$ is the lateral extent of the plates, $\alpha$ is the thermal expansion coefficient of the fluid, $\nu$ is its kinematic viscosity, and $\kappa$ is its thermal diffusivity.

Although Rayleigh-Bénard convection has been thoroughly studied, several important questions are not answered yet [35-37]. One of them is the heat transfer scaling for high Ra numbers. Several different scaling laws have been proposed, but none of them have been accepted as the definite answer. In order to study the scaling phenomena, the velocity fluctuations in both the boundary layer and bulk are analyzed [36]. As it will be shown below, anomalous diffusion manifests itself in the time series for velocity fluctuations in Rayleigh-Bénard convection, which suggests that the Hurst constant $H_1$ may be considered as a scaling parameter at small displacements.

Consider the data obtained in Ref. [36], which were kindly provided to us by the Delft University of Technology. The experiments were conducted in a 600 X 600 X 155 mm³ cell filled with water (a more detailed description of the experimental setup is presented in Ref. [35]), which corresponds to the aspect ratio of 4. Two copper plates at the top and bottom were kept at constant temperature by passing the flows of water differing in temperature through the plates' internal channels. The water was drawn from two thermostatically controlled (inaccuracy less than 0.03 K) basins. The plates thus imposed a controlled temperature difference on the working fluid. Velocities were measured by using two one-component laser Doppler anemometers manufactured by *Dantec*. One of the time series for the local fluctuations in the horizontal fluid velocity measured in the direction of the large-scale circulation in the bulk of the Rayleigh-Bénard cell (approximate sampling frequency $f_d$ = 3.7 Hz) is presented in Fig. 10(a). The time series was obtained at Ra = $4.2\times10^8$, Pr = 5.5, and Re = $4.2\times10^2$.

The results of FNS analysis of the signal shown in Fig. 10(a) are presented in Fig. 10(b)-(d), with the FNS parameters being given in the figure captions. It can be seen that the stochastic part $\Phi_s^{(2)}(\tau)$ of the experimental difference moment is adequately approximated by anomalous-diffusion interpolation (6).

In this case, $H_1 = 0.32$, which corresponds to "subdiffusion". It should be noted that the value of the diffusion coefficient $D$, which has the dimension of [m$^2$/s$^3$] in this example, listed in the caption of Fig. 10 has the physical meaning of the specific energy dissipation rate for the region where the local fluid velocity is measured. Therefore, this diffusion coefficient may be related to the Kolmogorov $\varepsilon$ parameter introduced in the theory of turbulence.

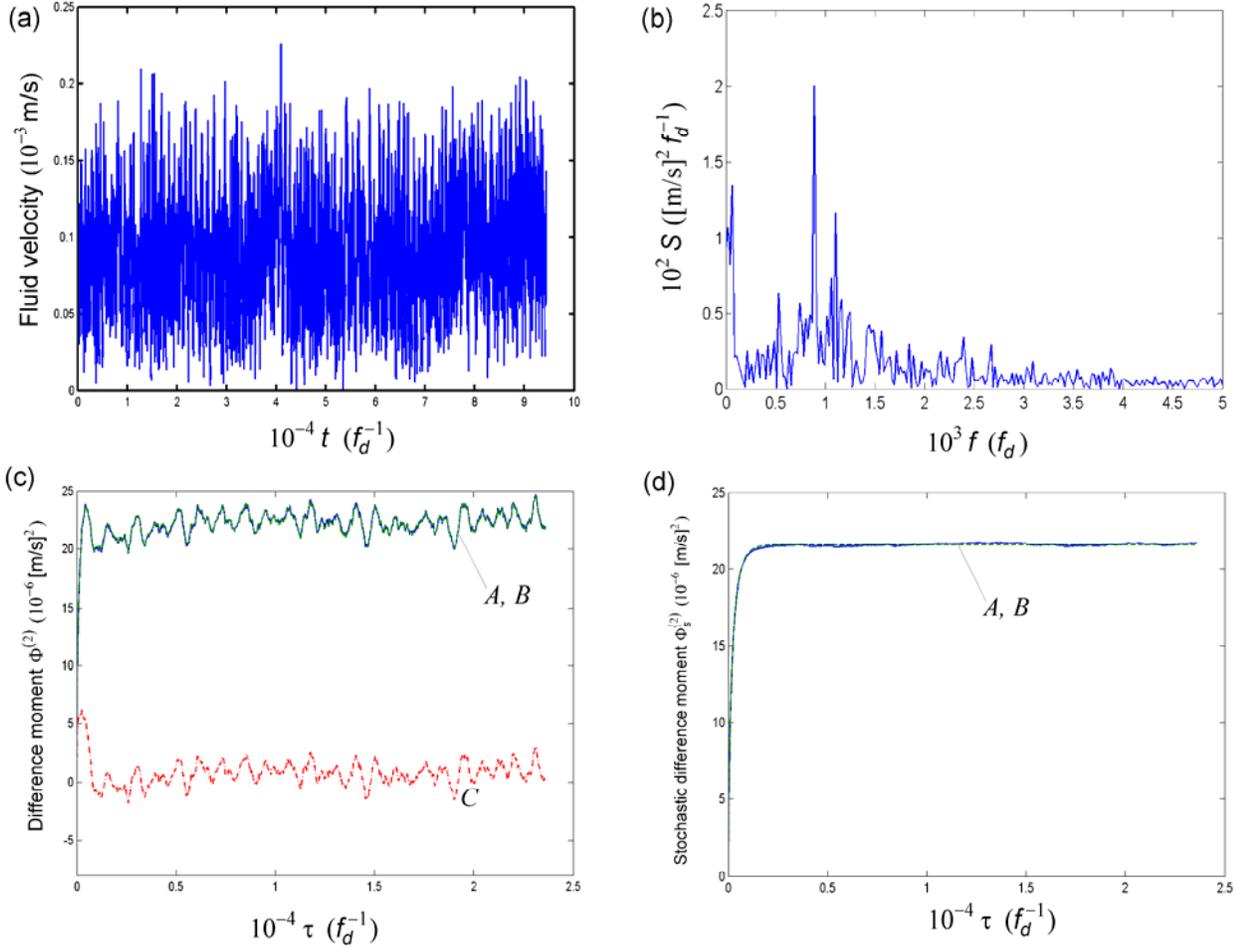

FIG. 10. (Color online) Analysis of the horizontal velocity component in the direction of the large-scale circulation in the bulk of the Rayleigh-Bénard cell [36] ($\sigma = 3.28 \times 10^{-3}$ m/s; $H_1 = 0.32$; $T_1 = 400\, f_d^{-1} = 108$ s, $D \approx 1.25 \times 10^{-7}$ m$^2$/s$^3$, $S_s(0) = 9.21 \times 10^{-3}$ (m/s)$^2 f^{-1}_d$; $n = 1.70$; $T_0 = 263\, f_d^{-1} \approx 71.1$ s): (a), source signal; (b), power spectrum $S(f)$ presented in Eq. (A.2) in the low-frequency range; (c), $A$ [solid blue line] – experimental $\Phi^{(2)}(\tau)$ presented in Eq. (A.8), $B$ [dashed green line] – general interpolation for $\Phi^{(2)}(\tau)$

given by the sum of expressions in Eqs. (A.7) and (A.10), C [dot-dashed red line] – $\Phi_r^{(2)}(\tau)$ presented in Eq. (A.7); (d), A [solid blue line] – $\Phi_s^{(2)}(\tau)$ presented in Eq. (A.9), B [dashed green line] – stochastic interpolation $\Phi_s^{(2)}(\tau)$ presented in Eq. (6).

**D. Anomalous diffusion in the geolectrical signals for seismic areas**

Geoelectrical parameters measured in seismic areas can help understand various complex phenomena related to seismic activity [38, 39]. For example, variations in the stress and fluid flow fields can produce changes in the geoelectrical field, resistivity, and other electrical parameters [40]. Therefore, the analysis of induced fluctuations may provide information on the governing mechanisms both in normal conditions and during intense seismic activity.

We will consider the time series of hourly self-potential measurements recorded in 2002 at station Giuliano located in a seismically active area in southern Italy [41]. Technically, a geoelectrical or self-potential time series is the sequence of voltage differences measured at a selected sampling interval using a receiving electrode array. During the geoelectrical soundings, when the electric current is introduced into the ground, the self-potential variations represent the noise. On the other hand, the main signal is measured using a passive measurement technique (i.e., without an energizing system). To avoid self-polarization effects, ceramic electrodes, which are ceramic vessels filled with a saturated solution of copper sulphate, were used.

Hourly time variations in the geoelectrical signal $V(t)$ measured at station Giuliano during January-September 2002 are illustrated in Fig. 11(a). The results of FNS analysis of the signals shown in Fig. 11(a) are presented in Fig. 11(b)-(d), with the FNS parameters being given in the figure captions. It can be seen that the stochastic part $\Phi_s^{(2)}(\tau)$ of the experimental difference moment is adequately approximated by anomalous-diffusion interpolation (6). In this case, $H_1 = 0.58$, which corresponds to the "superdiffusion" that is very close to the Fickian process.

The above analysis demonstrates that the stochastic changes in the geoelectrical signals measured in seismic areas may be described in terms of anomalous diffusion.

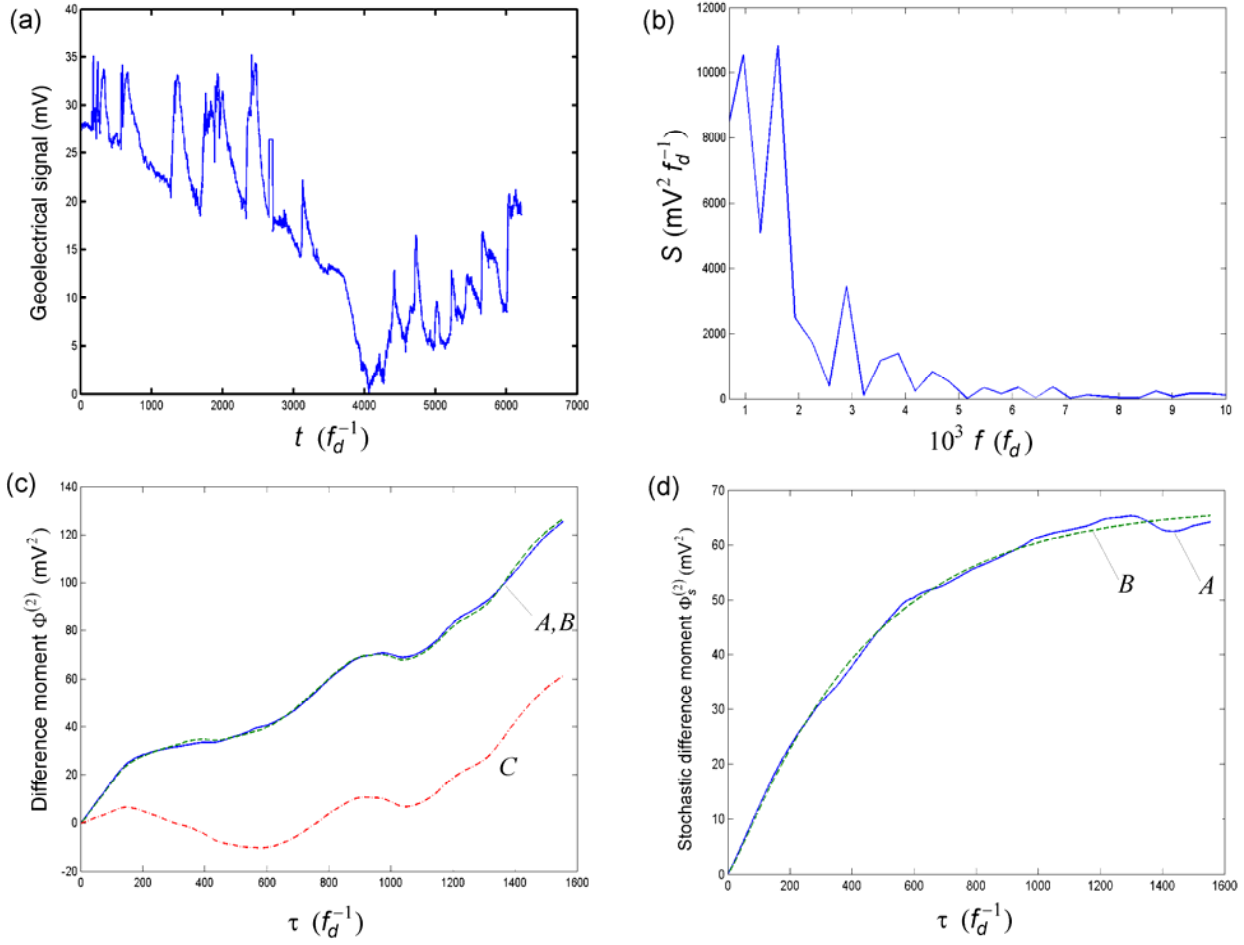

FIG. 11. (Color online) Analysis of hourly time variations in the geoelectrical signal measured at station Giuliano during January-September 2002 [41] ($\sigma = 5.80$ mV; $H_1 = 0.58$; $T_1 = 485\ f_d^{-1}$, $D \approx 0.087$ (mV)$^2$/hr, $S_s(0) = 6.58 \times 10^4$ (mV)$^2 f^{-1}_d$; $n = 2.11$; $T_0 = 452\ f_d^{-1}$): (a), source signal; (b), power spectrum $S(f)$ presented in Eq. (A.2) in the low-frequency range; (c), $A$ [solid blue line] – experimental $\Phi^{(2)}(\tau)$ presented in Eq. (A.8), $B$ [dashed green line] – general interpolation for $\Phi^{(2)}(\tau)$ given by the sum of expressions in Eqs. (A.7) and (A.10), $C$ [dot-dashed red line] – $\Phi_r^{(2)}(\tau)$ presented in Eq. (A.7); (d), $A$

[solid blue line] – $\Phi_s^{(2)}(\tau)$ presented in Eq. (A.9), *B* [dashed green line] – stochastic interpolation $\Phi_s^{(2)}(\tau)$ presented in Eq. (6).

## IV. SUMMARY AND CONCLUSIONS

The above analysis shows that interpolation expression (6) proposed in this study for the identification and parameterization of anomalous diffusion can be used both for small time intervals and when the stochastic variations of a dynamic variable reach a steady state after some characteristic interval. The interpolation expression represents a generalization of conventional formula (1), which is commonly used to determine the Hurst constant. Indeed, formula (1) is equivalent to asymptotic expression (7), which was derived from general interpolation (6) for small time intervals. In contrast to conventional formula (1), however, the ability of the interpolation expression to describe the mean-square-displacement function throughout the whole time interval makes it possible to determine not only the Hurst constant $H_1$ but also the standard deviation $\sigma$ and correlation loss time $T_1$, the characteristic time needed to reach the steady state. Consequently, interpolation formula (6) can immediately provide the value of diffusion coefficient by Eq. (26), which is a function of these three parameters.

It is seen in Figs. 4(c)-11(c) that real signals usually contain both regular and stochastic components. This requires the extraction of the stochastic component prior to estimating the parameters of anomalous diffusion from real signals. In this study, we used the parameterization procedure that includes the removal of regular (low-frequency) components from the second-order experimental difference moment built for the source time series and the fitting of the stochastic part of the experimental difference moment to the interpolation expression (described in detail in Appendix B).

The application of the parameterization procedure to the analysis of the stochastic dynamics for blinking fluorescence of quantum dots, X-ray emission from accreting objects, fluid velocity in Rayleigh-Bénard convection, and geoelectrical signals for seismic areas shows that the interpolation is able to adequately describe the stochastic part of the experimental difference moment for the signals presented in Figs. 4(a)-11(a). This implies that anomalous diffusion manifests itself in the stochastic dynamics for all four examples, which represent completely unrelated complex processes running at different scales: microscales for quantum dots and macroscales for stellar objects. The above suggests that anomalous diffusion can be identified using the proposed interpolation formula and parameterization algorithm in many other complex processes. One of the most promising potential applications for the parameterization procedure is the analysis of the data generated by single-molecule fluorescence spectroscopy in biological objects, which are characterized by the two-state kinetics of transitions between docked (active) and undocked (inactive) conformations [42, 43]. The time series of the observed photon emission in these systems are in many ways similar to the time series for blinking fluorescence of quantum dots studied in section III.A of this paper, which implies that anomalous diffusion can be identified and parameterized in this case as well. The values of the parameters of anomalous diffusion, including the diffusion coefficient and Hurst constant, may then be used to fine-tune and advance the mathematical models that describe the dynamics of state-to-state conformational transitions in biological structures, which are based on statistical methods and equations of chemical kinetics [44-48].

The above study shows that the interpolation expression proposed for the identification and parameterization of anomalous diffusion is practically equivalent in the particular case of Brownian diffusion to the mean square displacement calculated by the numerical solution to the problem given by the diffusion equation, initial impulse, and general integrodifferential symmetry boundary conditions. This makes it possible to model the stochastic dynamics of complex processes that can be described using the phenomenological interpolation expression. In the general case of anomalous diffusion, the mathematical models should include fractional-derivative differential equations with the integrodifferential boundary conditions incorporating the effects of nonstationarity and the finite residence times of the diffusion system in boundary "adstates".

**ACKNOWLEDGEMENTS**

The authors are grateful to Professor Masaro Kuno (University of Notre Dame) for the quantum-dot experimental data used for the analysis in section III, Joseph Neilsen (Harvard University, Department of Astronomy) for the discussion of the results presented in section IV, Jos Verdoold (Delft University of Technology) for the Rayleigh-Bénard experimental data used for the analysis in section V, and Professor Luciano Telesca (Istituto di Metodologie Avanzate di Analisi Ambientale, Consiglio Nazionale delle Ricerche, Area della Ricerca di Potenza) for the geoelectrical data used for the analysis in section VI.

This study was supported in part by the Russian Foundation for Basic Research, project no. 08-02-00230 *a*.

**APPENDIX A**

As both $S(f)$ and $\Phi^{(2)}(\tau)$ are defined in terms of the autocorrelation function, one may assume that these functions and the parameters characterizing them are tightly interrelated. As it will be shown below, however, the information contents of $S(f)$ and $\Phi^{(2)}(\tau)$ may be different, and in the general case we need the parameters for both functions to solve specific problems.

For simplicity, in this paper we will consider the problem of information difference between functions $S(f)$ and $\Phi^{(2)}(\tau)$ at a conceptual level. A more rigorous and substantiated analysis based on the theory of generalized functions [49] is presented elsewhere [15, 17]. Consider the process of one-dimensional "random walk" with low "kinematic viscosity" $\nu$ (Fig. A.1).

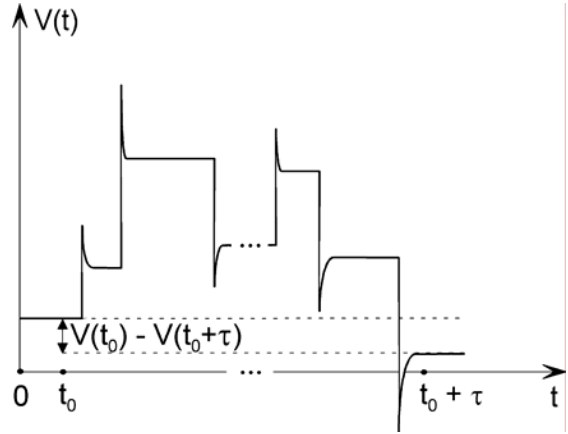

FIG. A.1. One-dimensional "random walk" with low "kinematic viscosity".

The small value of $\nu$ implies that when the signal passes from position $V_i$ to $V_{i+1}$, which are $|V_{i+1} - V_i|$ apart (in value) from each other, the system first overleaps ("overreacts") due to inertia and then "relaxes". We assume that the relaxation time is small compared to the residence time in the "fluctuation" position. It is obvious that when the number of walks is large, the functions $\Phi^{(2)}(\tau)$ will be independent of the values of "inertial overleaps" of the system and depend only on the algebraic sum of walk "jumps". At the same time, the functions $S(f)$, which characterize the "energy side" of the process, will depend on both spikes and jumps.

In this regard, note the well-known result illustrated by Fig. 51 in chapter 4.3 of Ref. [50], where an intermittent stochastic signal with alternating rapid stochastic spikes and laminar phases was considered. An artificial signal was generated by introducing a sequence of Dirac delta functions instead of rapid stochastic spikes. Then the power spectral density $S(f)$ for a sequence of $\delta$-functions with characteristic time intervals $T_0^i$ between adjacent $\delta$-functions on macroscopic time intervals $[-T/2, +T/2]$ ($T_0^i \ll T$) was calculated. It was shown that this artificial signal produced a flicker-noise function $S(f) \sim f^{-n}$ ($n \sim 1$) in the low-frequency spectrum range ($f \ll 1/2\pi T_0^i$). In other words, it was informative. On the other hand, if one would calculate the difference moment $\Phi^{(p)}(\tau)$ for this artificial signal (difference moment was not considered in Ref. [50]), it would be found that it is equal to zero because the domain set of a $\delta$-function sequence is a set of measure zero [49]. This statement can be easily illustrated by replacing $\delta$-functions in calculating $\Phi^{(p)}(\tau)$ with one of the well-known approximations; for example, using the Gaussian approximation with dispersion $\sigma_G^2$ and passing to the limit $\sigma_G \to 0$.

It should be emphasized that this separation of the information stored in various irregularities is attributed to the intermittent character of the evolution dynamics. Indeed, the information contents of $S_s(f)$ and $\Phi_s^{(2)}(\tau)$ coincide if there is no intermittence, as shown for the case of completely "irregular" dynamics of the Weierstrass – Mandelbrot (WM) function in Ref. [21]. Here, we use the subscript "s" for the functions $S_s(f)$ and $\Phi_s^{(2)}(\tau)$ to indicate the purely stochastic (no regular components) nature of the WM function.

It should be noted that although the real discrete signals measured at finite sampling frequency $f_d$ do not contain obvious spike (Dirac delta function) or jump (Heaviside function) irregularities, they usually show distinct stochastic "irregularities" characterized by different frequency ranges. In these cases, the lower and higher frequency ranges can be associated with jumps and spikes, respectively. As a result, the information contents of the power spectrum estimate and structural function are different, as was shown in Ref. [19]. When the sampling frequency is increased, the highest-frequency components begin to affect the values of the phenomenological parameters. The sampling interval itself in this case can be considered as an additional parameter that should be included in general parameterization procedures for real signals [19].

**APPENDIX B**

PARAMETERIZATION ALGORITHM (customized for MATLAB)

Notation

$q_{\min}$ is the number of frequency points to ignore when calculating $S_s(0)$ [step 3] and estimating resonant (regular) power spectrum [step 5];

$q_{max}$ is the highest frequency point used in calculating $S_s(0)$ [step 3];

Default values:

$$T_M = \frac{T}{4}, q_{min} = 0, q_{max} = q_{min} + 1.$$

STEP 1. Subtract the mean from the original signal.

STEP 2. Calculate the discrete cosine transform of the autocorrelator.

Calculate the autocorrelator:

$$\psi(m) = \frac{1}{N-m} \sum_{k=1}^{N-m} V(k)V(k+m) \quad \text{for } 0 \le m < M, \quad (A.1)$$

where

$T_M \le \frac{T}{2}$, $M = \left\lfloor \frac{T_M}{T} N \right\rfloor$ is the number of points on the frequency axis, and $N$ is the number of points in the averaging interval.

Fill in $\psi(m)$ for $m=M+1..2M-1$ using $\psi(m) = \psi(2M-m)$, as the autocorrelator is symmetric. In other words, we use only the first M+1 (0..M) values, which corresponds to our "interval of interest" $T_M$. Other values are obtained from the symmetry condition.

Calculate the fast Fourier transform of the autocorrelator:

$$S_{exp}(q) = \sum_{m=0}^{2M-1} \psi(m) \exp\left(-\frac{\pi i q m}{M}\right), \quad (A.2)$$

where $f$ is the frequency, $q = 2f T_M$, and $S(f) = \frac{S_{exp}(q)}{f_d}$.

For real signals, $S_{exp}(q) = S_{exp}(2M-q)$. So, keep only the first M+1 frequency points of $S_{exp}$.

For $q = 1..M-1$, multiply the values of $S_{exp}(q)$ by 2 (go to cosines from complex Fourier).

Result: $S_{exp}(q)$ for $q = 0..M$.

STEP 3. Calculate $S_s(0)$ using $S_{exp}$.

Select a specific frequency range $[q_{min}, q_{max}]$ in the range of low frequencies. Find the minimum value of $S_{exp}$ in the range and set $S_s(0)$ to this value. Formally,

$$S_s(0) = \min_{q \in [q_{min}, q_{max}]} S_{exp}. \tag{A.3}$$

STEP 4. Interpolate $S_{exp}(q)$ [$q>0$] with

$$S_s(q) \approx \frac{S_s(0)}{1+(2\pi \frac{q}{T_M} T_0)^n} \tag{A.4}$$

to find parameters $n$ and $T_0$.

Use the nonlinear method of least squares.

STEP 5. Calculate

$$S_r(q) = S_{exp}(q) - S_s(q). \tag{A.5}$$

For $q = 0..q_{min} - 1$, set $S_r(q) = 0$.

STEP 6. Calculate the autocorrelator for the resonant (regular) component.

For $q = 2..M - 1$, set $S_r = S_r/2$.

Use the symmetry complement to fill in $S_r$ for $m=M+1..2M-1$: $S_r(q) = S_r(2M - q)$.

Calculate the resonant autocorrelator

$$\psi_r(m) = \frac{1}{2M} \sum_{q=0}^{2M-1} S_r(q) \exp\left(\frac{\pi i q m}{M}\right). \tag{A.6}$$

Keep only the first $M+1$ points.

STEP 7. Calculate the difference moment for the resonant component:

$$\Phi_r^{(2)}(m) = 2[\psi_r(0) - \psi_r(m)] \text{ for } m = 0..M. \tag{A.7}$$

STEP 8. Calculate the difference moment for experimental series:

$$\Phi^{(2)}(m) = \frac{1}{N-m} \sum_{k=1}^{N-m} [V(k) - V(k+m)]^2 \text{ for } m = 0..M . \tag{A.8}$$

STEP 9. Calculate the difference moment for the stochastic component:

$$\Phi_s^{(2)}(m) = \Phi^{(2)}(m) - \Phi_r^{(2)}(m) \text{ for } m = 0..M . \tag{A.9}$$

STEP 10. Interpolate the stochastic difference moment using the function

$$\Phi_s^{(2)}(m) \approx 2\sigma^2 \cdot \left[ 1 - \Gamma^{-1}(H_1) \cdot \Gamma\left(H_1, \frac{m}{f_d T_1}\right) \right]^2 , \tag{A.10}$$

$$\Gamma(s,x) = \int_x^\infty \exp(-t) \cdot t^{s-1} dt, \ \Gamma(s) = \Gamma(s,0) .$$

Find parameters $\sigma$, $H_1$, $T_1$.

STEP 11. Calculate the jump component of the power spectrum

$$S_{sJ}(q) \approx \frac{S_{cJ}(0)}{1 + (2\pi \frac{q}{T_M} T_1)^{2H_1+1}} , \tag{A.11}$$

where

$$S_{sJ}(0) = 4\sigma^2 T_1 H_1 \cdot \left\{ 1 - \frac{1}{2H_1 \Gamma^2(H_1)} \int_0^\infty \Gamma^2(H_1,\xi) d\xi \right\} .$$